\documentclass[twocolumn,prl,aps,showpacs]{revtex4}
\usepackage{amssymb}

\usepackage{amsmath}
\usepackage{graphicx}
\usepackage{dcolumn}
\usepackage{bm}

\input{tcilatex}

\begin{document}

\title{A neutron scattering study of two-magnon states in the quantum magnet
copper nitrate}
\author{D.\ A. Tennant$^{1,2,3,4}$, C. Broholm$^{5,6}$, D. H. Reich$^{7}$,
S.\ E. Nagler$^{3}$, G.\ E. Granroth$^{3}$, {T.~Barnes}$^{7,8}$, K. Damle$%
^{9}$, G. Xu$^{5}$, Y. Chen$^{5}$, and B.\ C. Sales$^{3}$.}
\affiliation{$^{1}$ISIS Facility, Rutherford Appleton Laboratory, Didcot OX11 0QX, Oxon.,
United Kingdom}
\affiliation{$^2$Oxford Physics, Clarendon Laboratory , Parks Road, Oxford OX1 3PU,
United Kingdom}
\affiliation{$^{3}$Solid State Division, Oak Ridge National Laboratory, Oak Ridge, TN
37831-6393\\
$^{4}$Dept. of Physics and Chemistry, Ris\o\ National Laboratory, Roskilde,
DK-4000, Denmark}
\affiliation{$^{5}$Department of Physics and Astronomy, The Johns Hopkins University,
Baltimore, MD 21218}
\affiliation{$^{6}$NIST Center for Neutron Research, National Institute of Standards and
Technology, Gaithersburg, MD 20899 \\
$^{7}$Department of Physics and Astronomy, University of Tennessee,
Knoxville, TN 37996-1501}
\affiliation{$^{8}$Physics Division, Oak Ridge National Laboratory, Oak Ridge, TN
37831-6373\\
$^{9}$Physics Department, Harvard University, Cambridge, MA 02138}
\date{\today}

\begin{abstract}
We report measurements of the two-magnon states in a dimerized
antiferromagnetic chain material, copper nitrate ({Cu(NO$_{3}$)$_{2}\cdot
2.5 $D$_{2}$O}). Using inelastic neutron scattering we have measured the one
and two magnon excitation spectra in a large single crystal. The data are in
excellent agreement with a perturbative expansion of the alternating
Heisenberg Hamiltonian from the strongly dimerized limit. The expansion
predicts a two-magnon bound state for $q\sim (2n+1)\pi d$ which is
consistent with the neutron scattering data.
\end{abstract}

\maketitle


\section{Introduction}

The basic physics of the elementary one-magnon excitations of
lower-dimensional quantum antiferromagnets can now be regarded as well
established, both theoretically and experimentally through studies of
materials that are reasonably accurate realizations of the spin
Hamiltonians. In contrast, higher excitations such as multimagnon continua
and bound states have attracted relatively little attention. This topic may
prove to be a fascinating area for the application of few- and many-body
techniques, and will involve interesting and nontrivial results in band
structure, band mixing, bound state formation, phase transitions through the
formation of condensates of magnetic excitations \cite%
{Oosawa01,nikuniprl2000,
WatsonMeisel01,YuHaas00,TsvelikGiamarchi,Affleck91,coldea02}, and other
collective phenomena.

Although some aspects of the physics of low-lying multimagnon states can be
inferred from model Hamiltonians using standard theoretical techniques, few
experimental studies of these higher excitations have been reported to date.
Experimental difficulties that have precluded such work include relatively
weak couplings of probes to these higher excitations, dominant contributions
from the lowest one-magnon excitations, and resolution requirements in
energy and wavenumber that are beyond the capabilities of most techniques.

High-resolution inelastic neutron scattering should prove to be an ideal
technique for observing some of these higher magnetic excitations. One can
control both energy and momentum transfer, so that the existence and
spectral weight of higher magnetic excitations can be established and
quantified. The new generation of high-intensity neutron sources combined
with high-resolution detectors should allow the observation of details of
the multimagnon excitation spectrum such as band boundaries, quantitative
determination of the dynamical correlation function $\mathcal{S}(\mathbf{Q}%
,\omega )$ and discontinuities within a band \cite{barnes02}, and weakly
bound states just below the band edge. The principle limitation in this
approach may be the unavoidable $\Delta S=1$ selection rule of magnetic
neutron scattering, so that one can only reach spin $S=1$ excitations given
an $S=0$ ground state and an isotropic spin Hamiltonian. Other techniques
such as Raman scattering can be used to study certain of these higher
excitations, \textit{albeit} with strong constraints on the accessible spin
and momentum quantum numbers.

Systems that appear especially interesting for studies of higher magnetic
excitations at present are quasi-1D spin chains and spin ladders, since many
of these have gaps and hence will have separated bands of higher excitations
and perhaps bound states of magnons. The alternating Heisenberg
antiferromagnetic chain (AHC) with spin-1/2 is an example of such a system;
with any amount of alternation $0<\alpha <1$ (where $\alpha \equiv J2/J1$)
the AHC has an energy gap in its one-magnon dispersion with a second gap to
the multimagnon continuum, and two-magnon bound states with spin-0 and
spin-1 are predicted \cite{uhrig}. The AHC is also attractive because of its
relative simplicity and because of recent extensive theoretical studies of
the low-energy excitations in this model.

In this paper we present results from an inelastic neutron scattering study
of higher magnetic excitations in copper nitrate ({Cu(NO$_{3}$)$_{2}\cdot
2.5 $D$_{2}$O}), which was recently confirmed by neutron scattering \cite%
{xu00} to be an accurate realization of a strongly alternating Heisenberg
antiferromagnetic chain. This material is especially attractive because the
alternation parameter $\alpha \approx 0.27$ is close to the value predicted
to maximize the separation of the spin-1 two-magnon bound state from the
continuum \cite{BRT}. In addition it is relatively easy to prepare large
single crystals of this material, which compensates for the weak neutron
scattering intensity from the higher magnetic excitations.
\begin{figure}[ht]
\includegraphics[viewport=0 120 500 600,width=3in,clip]
{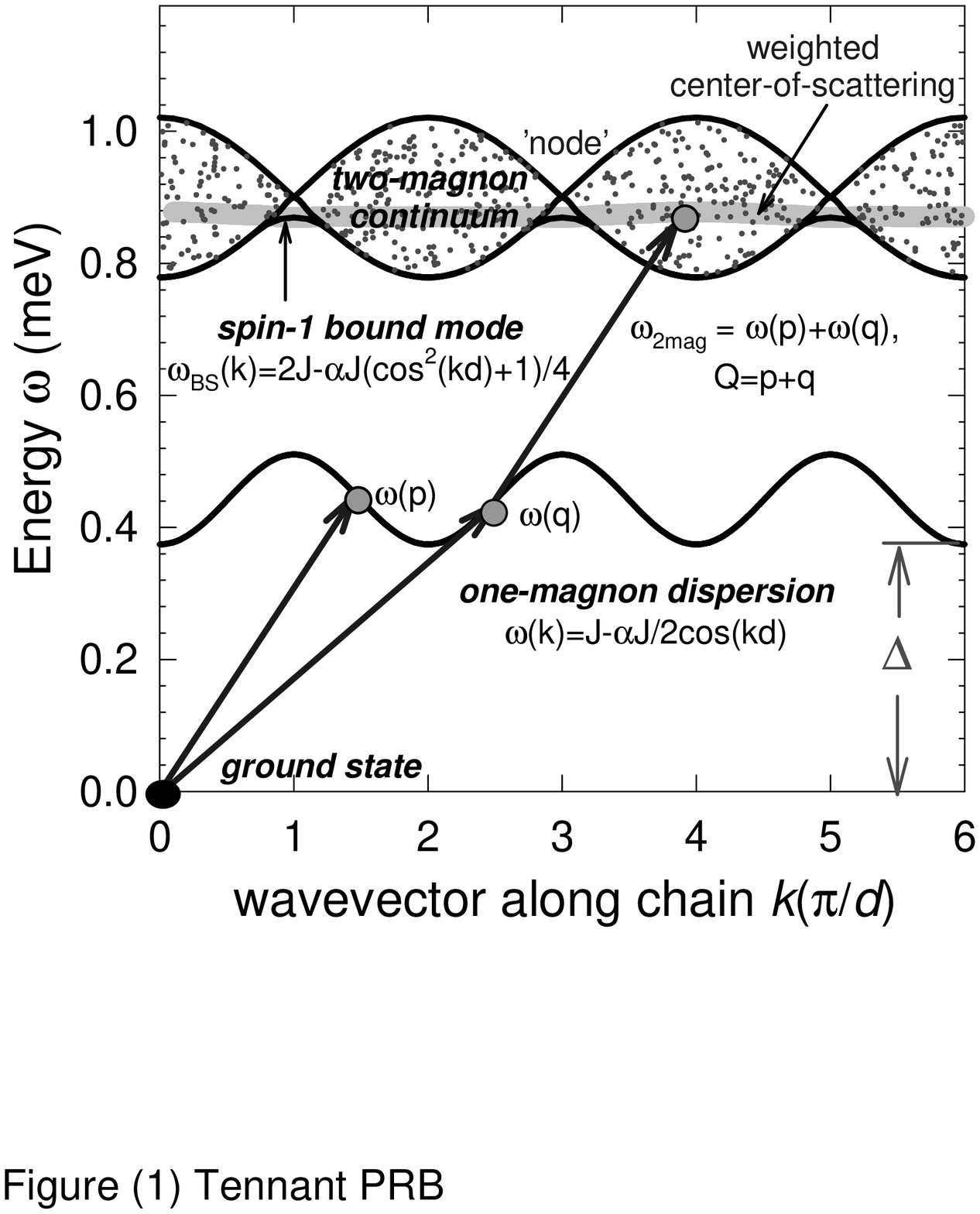}
\caption{\label{fig:epsart} The $S=1$ excitation spectrum of the spin-1/2
alternating Heisenberg chain with moderately strong dimerization. The
parameters used are $J=0.45$ meV and $\protect\alpha =0.27$, (as fit to
copper nitrate neutron scattering data.) There is an energy gap $\Delta $
from the $S=0$ ground state to the $S=1$ triplet magnon band. The two-magnon
continuum states are also shown; these have energies and wavevectors given
by the sum of two independent one-magnon excitations. A $S=1$ bound state is
predicted to lie just below the two-magnon continuum for small wavevectors
near $k=\protect\pi /d$, where the continuum has minimum width.}
\end{figure}

The paper is arranged as follows: Section (2) summarizes important results
from the theory of the ground and excited states of the alternating
Heisenberg chain. Section (3) reviews the magnetic properties of {Cu(NO$_{3}$%
)$_{2}\cdot 2.5$D$_{2}$O}, or CN for short. Section (4) presents the results
of our measurements, with the analysis in terms of the model given in
section (5). A discussion of the results is given in (6), with conclusions
in (7). Additional theoretical results for two-magnon excitations in this
model are given in the Appendix.

\section{Theory}

The spin-1/2 alternating Heisenberg chain has received much attention in the
theoretical literature. This simple model plays a central role in the study
of the spin-Peierls effect, and is also known to provide an accurate
description of the magnetic properties of many real materials. The model
consists of antiferromagnetically coupled Heisenberg spin pairs, ``dimers'',
which are themselves coupled by weaker antiferromagnetic Heisenberg
interactions in an alternating chain, as shown in Fig.2. The Hamiltonian for
this model is given by 
\begin{equation}
H=\sum_{m=1}^{N/2}\ J\;\bigg\{\mathbf{S}{_{m,-}}\cdot \mathbf{S}{_{m,+}}%
+\alpha \;\mathbf{S}{_{m,+}}\cdot \mathbf{S}{_{m+1,-}}\bigg\}\ ,
\end{equation}%
where $N$ is the number of spins in the chain, $J>0$ (also called $J_{1}$)
is the intradimer coupling, $\alpha J$ (also $J_{2}$) is the interdimer
coupling, and $\alpha $ is allowed the range $0<\alpha <1$. The index $m$
labels the dimers, and $-$ and $+$ denote left and right spins. The position
of each spin is given by $\mathbf{r}{_{m,\pm }=}m\mathbf{d}\pm \mathbf{\rho }%
/2$, where $\mathbf{d}$ is the chain repeat vector, and $\mathbf{\rho }$ is
the intradimer separation.
\begin{figure}[ht]
\includegraphics[viewport=0 280 500 520,width=3in,clip]
{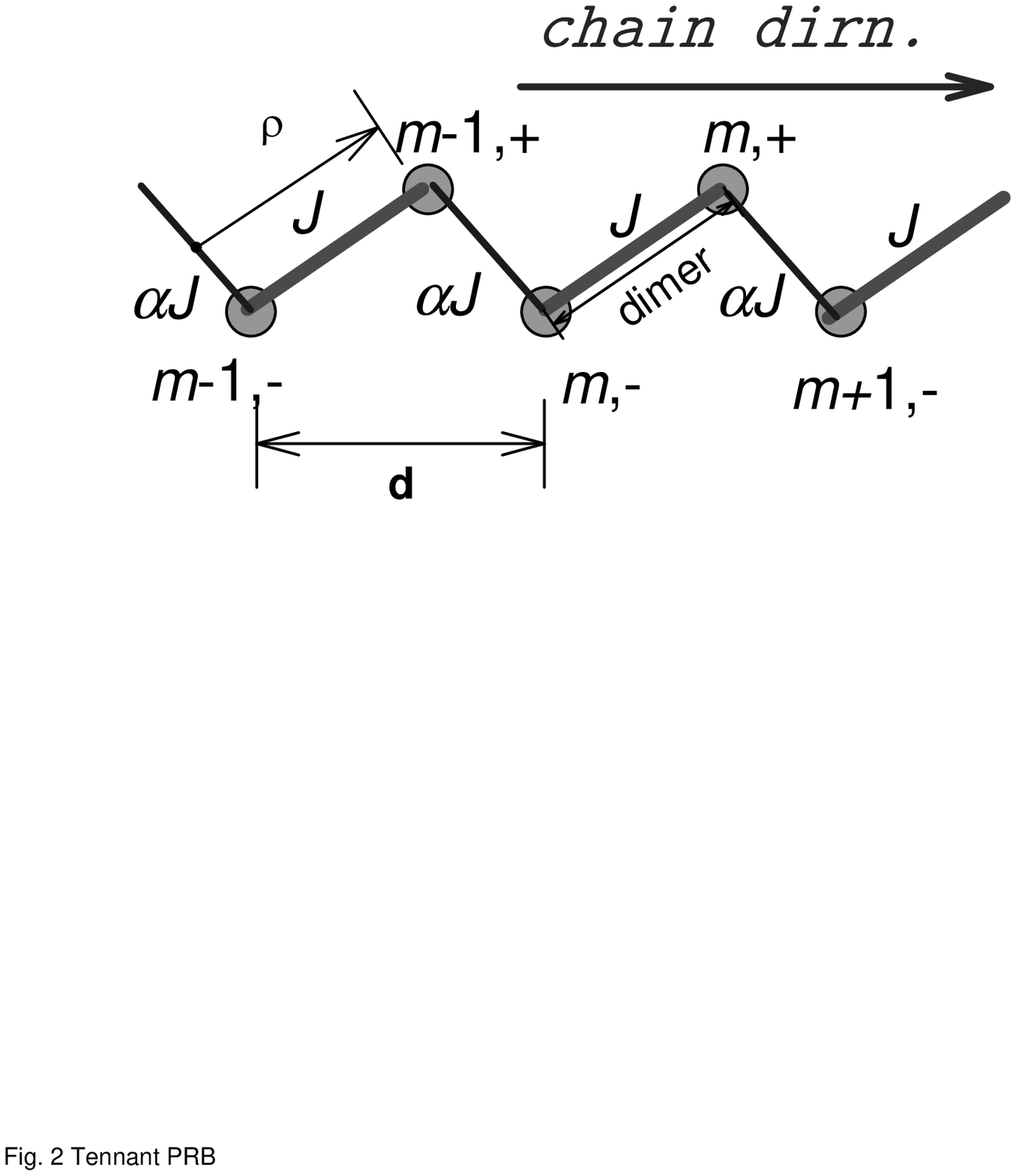}
\caption{\label{fig:num2}Alternating chain layout showing dimers coupled together. Each dimer is
labeled by an index $m$, $\mathbf{\protect\rho }$ is the separation between
dimer spins, and $\mathbf{d}$ is the chain repeat vector.}
\end{figure}

This model has a nontrivial spin-0 ground state, and for all $\alpha $ in
the allowed range $0<\alpha <1$ has a gap to the first excitation, which is
a band of spin-1 excitations (magnons). In the ``strong-coupling limit'' $%
\alpha \ll 1$ the ground state approaches a system of uncoupled spin-0
dimers, and the one-magnon excitations can be accurately described as a
single dimer excited to spin-1 (an $``$exciton''), delocalized on the chain
to give states of definite along-chain wavenumber $k$. The energies and some
matrix elements of these states have been evaluated as power series in the
intradimer coupling $\alpha $ \cite{BRT,CCM}.

\begin{table}[tbp]
\caption{Eigenstates of dimer $m$, $H_{m}=J\mathbf{S}_{m,-}\cdot \mathbf{S}%
_{m,+}$. }%
\begin{ruledtabular}
\begin{tabular}{ccccc}
Label&$\Psi_m$&$E$&$S^{z}_m$&$S_m$\\
\hline
$G_m$ & $\frac{1}{\sqrt{2}}\left\{ |\uparrow \downarrow \rangle
-|\downarrow \uparrow \rangle \right\} $ & $-3J/4$ & $0$ & $0$ \\ 
$1_m$ & $|\uparrow \uparrow \rangle $ & $J/4$ & $1$ & $1$
\\
$0_m$ & $\frac{1}{\sqrt{2}}\left\{ |\uparrow \downarrow \rangle
+|\downarrow \uparrow \rangle \right\} $ & $J/4$ & $0$ & $1$ \\ 
$\overline{1}_m$ & $|\downarrow \downarrow \rangle $ & $J/4$ & $-1$ & $1$ \\
\end{tabular}
\end{ruledtabular}
\end{table}

To interpret experimental results approximate analytic forms of the
wavefunctions and energies are useful and we include calculations of
wavefunctions expanded around the single dimer eigenstates in the Appendix.
The one-magnon $S=1$ triplet has a gap energy $\Delta =J-\alpha J/2$ and
dispersion to $\mathcal{O}(\alpha )$%
\begin{equation}
\omega _{\text{1mag}}(k)=J-\alpha J/2\cos (kd).
\end{equation}%
This magnon wavefunction can be visualized as a localized wavepacket of
magnetic polarization (excitons) along $x,y,$ or $z$ carrying total spin-1
travelling through a featureless singlet background with the gap energy
coming from the effort expended in breaking a dimer bond.

Approximate wavefunctions for the two-magnon states are given in the
Appendix. At large separations the magnons do not overlap and they behave as
free particles, however when close they interfere and scatter off each
other. In a one dimensional geometry such scattering conserves particle
number, energy, and momentum up to a lattice wavevector, and the scattering
introduces a momentum dependent phase shift in the scattered wavefunction.
The energy of these states (to order $\mathcal{O}(1/N)$) is given by $\omega
_{k_{1},k_{2}}(k)=\omega _{\text{1mag}}(k_{1})+\omega _{\text{1mag}}(k_{2})$%
, where $k=k_{1}+k_{2}$ is the total wavevector. The resulting continuum
using equation (2) is illustrated in Fig. 1 \cite{barnes02}.

As well as elastic scattering of magnons, bound states also form. To
appreciate their physical origin consider two dimers, both in excited
states, coupled by a single interdimer coupling $\alpha J$. In the absence
of coupling all the double excited states $S=0,1,$ and $2$\ have the same
energy $2J$. However, the interdimer coupling splits these states. There is
an $S=2$ quintuplet of energy $2J+\alpha J/4$ (this energy is higher because
the interdimer coupling favours antiferromagnetism whereas the $S=2$\ states
have all spins along the same direction - ferromagnetic). There is also an $%
S=1$ triplet of energy $2J-\alpha J/4$, this lowering of energy is purely
quantum mechanical and comes from resonance between the two excited dimers.
Finally there is an $S=0$ singlet of even lower energy, $2J-\alpha J/2$,
which gains resonance and also antiferromagnetic energy due to the spins in
neighboring dimers pointing in opposite directions.

As the $S=0$ and $1$ states with excited dimers neighboring each other have
lower energy than two well separated excited dimers (by $\alpha J/4$ and $%
\alpha J/2$ respectively) there is a short range attractive potential.
Magnons can be confined within the potential well (bound states) as long as
the relative kinetic energy between the magnons is smaller than the
interaction energy.

This situation applies to the bound states in the AHC, the $S=0$ mode at all
wavevectors and the $S=1$ mode over limited wavevectors around the node
points. These are positions where the kinetic energy is small compared to
the binding potential. The $S=1$ bound state is visible to neutron
scattering and has a dispersion \cite{uhrig}\cite{Damle} 
\begin{equation*}
\omega _{\text{BS}}=2J-\frac{\alpha J}{4}\left( 4\cos ^{2}(kd/2)+1\right) .
\end{equation*}%
The bound state is characterized by a probability amplitude for the
separation between the two magnons that drops exponentially with distance
(see appendix for more details). These $S=1$ bound states exist only over
the range $\left| n\pi -kd\right| \leqslant \pi /3$, where $n$ is an odd
integer, see Fig. 1.

\section{Magnetic properties of {Cu(NO$_{3}$)$_{2}\cdot 2.5$D$_{2}$O}}

The structural and magnetic properties of CN have been thoroughly
investigated and shown to be near ideal. The structure of {Cu(NO$_{3}$)$%
_{2}\cdot 2.5$H$_{2}$O was investigated by Garaj \cite{garaj} and Morosin %
\cite{morosin}, and was shown to have a monoclinic crystal structure with
space group $I12/c1$ \cite{space_group} and the lattice parameters and
crystallographic copper positions given in Table II.\ The deuterated form {%
Cu(NO$_{3}$)$_{2}\cdot 2.5$D$_{2}$O}\ we studied has low temperature ($T=3$
K) lattice parameters }$a${$=16.1$, }$b${$=4.9$, $c=15.8$~\AA\ and $\beta
=92.9^{\circ }$\cite{xu00}. }

The magnetism of CN\ arises from the Cu$^{2+}$ ions. Crystal electric fields
from the oxygen ligands surrounding the Cu$^{2+}$ ions quench their orbital
moments, leaving a near-isotropic spin-1/2 moment with g-values that show a
small easy-axis anisotropy along the crystallographic $b$-direction; the
values are $g_{\parallel \mathbf{b}}=2.33$ and $g_{\perp \mathbf{b}}=2.09$ %
\cite{diederix_paper1}. Magnetic superexchange in this material is mediated
by long double Cu-O-O-Cu exchange paths, which accounts for the rather weak
exchange interaction observed in CN.

\begin{table}[tbp]
\caption{Crystallographic data for {Cu(NO$_{3}$)$_{2}\cdot 2.5$H$_{2}$O}
from reference \protect\cite{morosin}. The room temperature lattice
parameters are $a=16.453$, $b=4.936$, $c=15.963$~\AA\ and $\protect\beta %
=93.765^{\circ }$ The Cu ions are at the 8f positions at (x,y,z) where
x=0.12613, y=0.01352, z=0.11376. The equivalent positions in the unit cell
are:}%
\begin{ruledtabular}
\begin{tabular}{cccc}
Atom&$x$-pos&$y$-pos&$z$-pos\\
\hline
$1$ & $x$ & $y$ & $z$ \\ 
$2$ & $1-x$ & $1-y$ & $1-z$ \\ 
$3$ & $1-x$ & $y$ & $\frac{1}{2}-z$ \\ 
$4$ & $x$ & $1-y$ & $\frac{1}{2}+z$ \\ 
$5$ & $\frac{1}{2}+{x}$ & $\frac{1}{2}-y$ & $z$ \\ 
$6$ & $\frac{1}{2}-{x}$ & $\frac{1}{2}+y$ & $1-z$ \\ 
$7$ & $\frac{1}{2}-{x}$ & $\frac{1}{2}-y$ & $\frac{1}{2}-z$ \\
$8$ & $\frac{1}{2}+{x}$ & $\frac{1}{2}+y$ & $\frac{1}{2}+z$ \\ 
\end{tabular}
\end{ruledtabular}
\end{table}

The exchange couplings of CN were studied by Eckert \textit{et al.} in \cite%
{eckert79} and are illustrated in Fig. 3. The dominant magnetic exchange
integral $J$ is between pairs of spins (copper positions 7 \&\ 8, and
equivalent pairs, \textit{c.f. }Fig. 3) forming dimers with a separation of
5.3 \AA . These dimers separated by \emph{crystal vector} $\mathbf{u}=\left[
u_{1},u_{2},u_{3}\right] \equiv u_{1}\mathbf{a}+u_{2}\mathbf{b}+u_{3}\mathbf{%
c}$ are coupled together by exchanges $J_{\mathbf{u}}^{\prime }$; the only
exchange paths of appreciable strength are $J_{[{\frac{1}{2}},\pm {\frac{1}{2%
}},{\frac{1}{2}}]}^{\prime }$, connected via bonds between 1 \&\ 7 and
equivalent (bond length 6.2 \AA ). This results in two sets of $S=1/2$
alternating Heisenberg chains running in the $[1,1,1]$ and $[1,\overline{1}%
,1]$ directions of the crystal with repetition every $\mathbf{d}=[1,1,1]/2$
and $\mathbf{d}^{\prime }=[1,\overline{1},1]/2$\ respectively (repeat
distance $d=11.3$ {\AA )}; the corresponding intradimer vectors are $\mathbf{%
\rho }=[0.252,\pm 0.027,0.228]$ (where the $x,y,z$ positions from Table II\
have been used). Inelastic neutron scattering measurements \cite{xu00} have
recently further confirmed the model alternating Heisenberg chain properties
of CN and have shown that the dominant collective excitations are indeed the
gapped triplet of magnons expected for the AHC.
\begin{figure}[ht]
\includegraphics[viewport=20 20 400 370,width=3in,clip]
{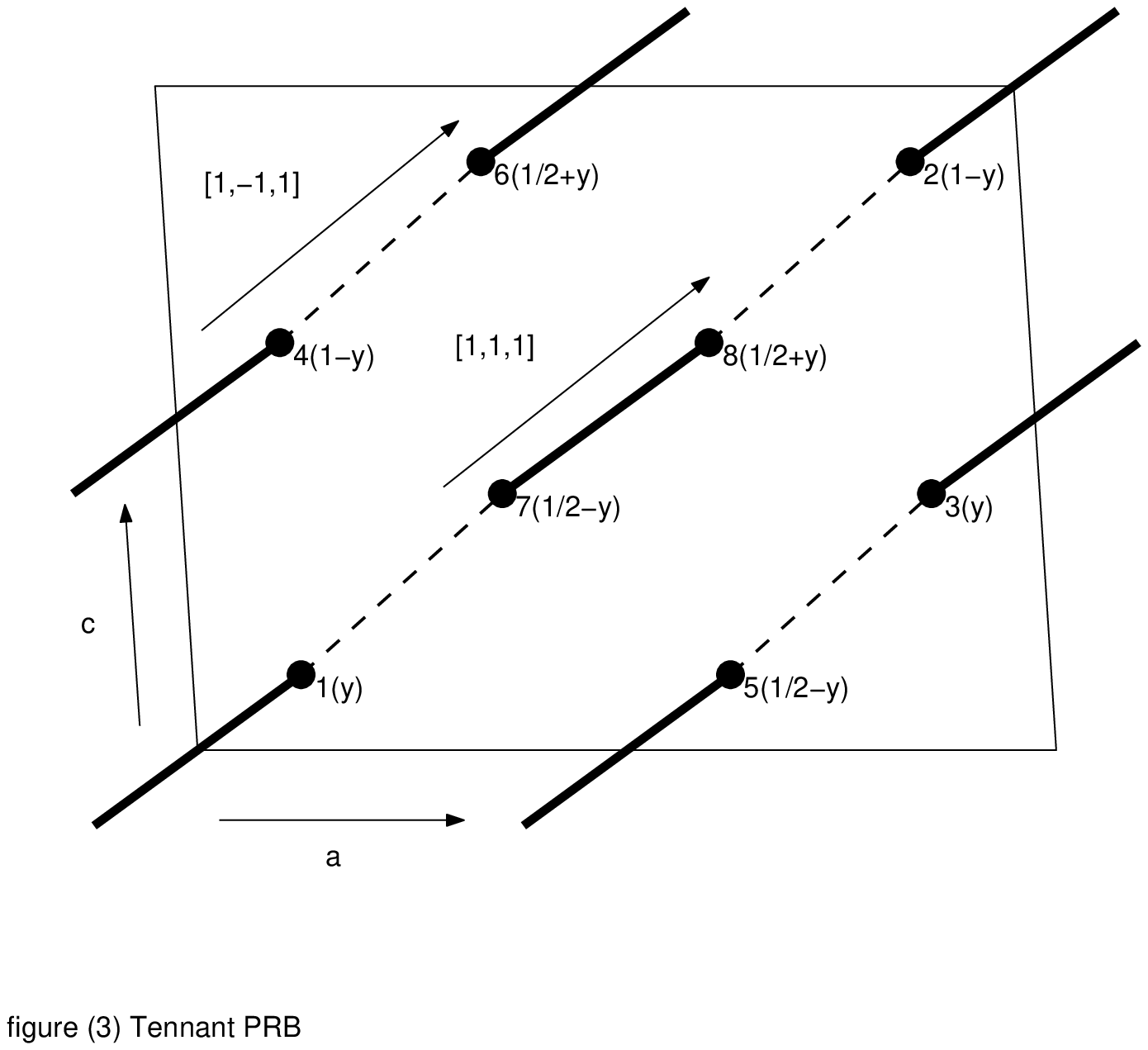}
\caption{\label{fig:num3} Positions of copper ions projected onto the $ac$ plane
for Cu(NO$_{3}$)$_{2}\cdot 2.5$D$_{2}$O. The atom positions are those
detailed in Table II. Two sets of identical chains run in the $\left[ 1,1,1%
\right] $ and $\left[ 1,\overline{1},1\right] $ directions respectively.}
\end{figure}

\subsection{Approximate ground state energy}

Bulk magnetic measurements give information on the gap, exchange, and ground
state energies of the spin chains in copper nitrate. Measurements of the
effect of applied magnetic field show that spin flop (SF) ordering is
induced in CN above a critical field $B_{c1}\approx 2.7$~T, with a
transition to full alignment at $B_{c2}\approx 4.3$~T and there are no
demagnetization effects in the zero temperature limit \cite{Diederix}. As
the orbital moment on the Cu$^{2+}$ ions is quenched by the crystal electric
field and demagnetization effects are negligible, the field $B_{c1}$ can be
used to directly give the excitation gap energy to the one-magnon states,
and Diederix \textit{et al.} report a value of $\Delta =0.378\pm .007$~meV %
\cite{Diederix}.

The gap can be inferred from the transition field because the magnons carry
spin quantum numbers $S^{z}=1,0,-1$ and are split into three dispersive
modes shifted by a Zeeman energy with respect to each other. The ground
state by virtue of its spin-0 quantum number is unaffected by the field and
the transition occurs when the Zeeman energy of the lowest mode closes the
gap and magnons condense into the ground state. The long range order itself
is due to weak couplings between the chains.

The high field transition, $B_{c2}$, yields further important information.
It is where all the low-lying magnons are completely condensed into the
ground state and the spins are fully aligned along the field. The fully
aligned state is an exact eigenstate, and for an unfrustrated quantum magnet
the transition field gives the sum of exchange couplings $g\mu
_{B}B_{c2}=J+\sum_{\mathbf{u}}J_{\mathbf{u}}^{\prime }=0.580\pm .007$~meV in
the system \cite{coldea02}. An estimate of the ground state energy of CN can
be made using these numbers.

Using the low temperature isothermal magnetization $M(B)=g\mu
_{B}\left\langle S^{z}\right\rangle _{B}$ as a measure of the work required
to saturate the spin chains from the zero-field quantum ground state, an
energy-per-spin can be inferred. The zero-field ground state energy-per-spin 
$e_{0}$ can be estimated via the formula $e_{0}\approx e_{f}-g\mu
_{B}SB_{c2}+\int_{0}^{B_{c2}}M(B)dB$ where the fully-aligned energy-per-spin
is $e_{f}=S^{2}/2\cdot (J+\sum_{\mathbf{u}}J_{\mathbf{u}}^{\prime })=g\mu
_{B}B_{c2}/8=0.0725\pm .001$~meV for a $S=1/2$ unfrustrated system.
Utilizing the 270 mK data of Diederix \textit{et al.} in Fig. 3 of \cite%
{Diederix} (measured using proton resonance) to determine the integral over
magnetization gives an experimental ground state energy-per-spin $%
e_{0}=-0.174\pm .004$~meV. This is essentially the $T=0$ result, as the gap
activation energy corresponds to 4.4 K.

To estimate thermodynamic properties we approximate the sum of the \emph{%
inter}dimer\ exchanges by the single coupling $\alpha J=\sum_{\mathbf{u}}J_{%
\mathbf{u}}^{\prime }$ of equation (1). Using the $\mathcal{O}(\alpha ^{9})$
expansions \cite{BRT} for $\Delta (\alpha )$ and $e_{0}$ gives $J=0.455\pm
.002$~meV and $\alpha =0.277\pm .006$; in agreement with the results of \cite%
{Diederix} and \cite{Bonner}, $J=0.45$ meV and $\alpha =0.27$. Our
calculated values of the thermodynamic parameters $J+\sum_{\mathbf{u}}J_{%
\mathbf{u}}^{\prime }=0.581$ meV, $\Delta =0.379$ meV and $e_{0}=-0.172$~meV
agree within error with the experimental values.

\section{Experimental Method}

\subsection{Neutron scattering}

The inelastic neutron-scattering cross-section \cite{marshall71}%
\begin{eqnarray*}
\frac{d^{2}\sigma }{d\Omega d\omega } &\propto &N\sigma
_{mag}\sum\limits_{\alpha ,\beta }\frac{k_{f}}{k_{i}}\left| F(\mathbf{Q}%
)\right| ^{2} \\
&&\left( \delta _{\alpha \beta }-Q_{\alpha }Q_{\beta }\right) \mathcal{S}%
^{\alpha \beta }(\mathbf{Q},\omega ),
\end{eqnarray*}%
is proportional to the dynamical response $\mathcal{S}^{\alpha \beta }(%
\mathbf{Q},\omega )$, where $\mathbf{Q}$ is the wavevector transfer, $%
F\left( \mathbf{Q}\right) $ is the magnetic form factor, $N$ is the number
of scattering centers, the constant $\sigma _{mag}=0.2896$ b,\ $k_{i}$\ and $%
k_{f}$\ are the momenta of initial and final neutron states respectively, $g$
is the Land\'{e} $g$-factor,\ and $\alpha =x,y,z$ are Cartesian coordinates.
The dynamical response is the space and time Fourier transform of the
spin-spin correlation function 
\begin{eqnarray*}
\mathcal{S}^{\alpha \beta }(\mathbf{Q},\omega ) &=&\frac{1}{2\pi N}%
\sum_{i,j}\int \exp \left( i\left( \omega t+\mathbf{Q}\cdot (\mathbf{r}_{i}-%
\mathbf{r}_{j}\right) \right) \\
&&\langle {S}_{i}^{\alpha }(0)S_{j}^{\beta }(t)\rangle dt,
\end{eqnarray*}%
where $i$ and $j$ labels sites of the system. For the AHC, equation (1),
spin conservation and isotropy in spin space ensure that $\mathcal{S}%
^{\alpha \beta }(\mathbf{Q},\omega )=0$ for $\alpha \neq \beta $, and all
diagonal spin components are equivalent $\mathcal{S}^{xx}(\mathbf{Q},\omega
)=\mathcal{S}^{yy}(\mathbf{Q},\omega )=\mathcal{S}^{zz}(\mathbf{Q},\omega ).$
At $T=0$ this is given by 
\begin{equation*}
\mathcal{S}^{xx}(\mathbf{Q},\omega )=\frac{1}{2}\mathcal{S}^{+-}(\mathbf{Q}%
,\omega )=\frac{1}{2}\sum_{\lambda }|\langle {\Psi _{\lambda }}(k)|S_{%
\mathbf{Q}}^{+}|\Psi _{G}\rangle |^{2}\delta (\omega -\omega _{\lambda }),
\end{equation*}%
where $\lambda $ label the eigenstates of $H$ and%
\begin{equation*}
S_{\mathbf{Q}}^{+}=\frac{1}{\sqrt{2N_{d}}}\sum_{m=1}^{N_{d}}\sum_{p=\pm
}\exp (i\mathbf{Q}\cdot \mathbf{r}_{m,p})S_{m,p}^{+}
\end{equation*}%
is the Fourier transformed spin creation operator.

The action of the neutron is to flip a spin and so create a localized spin-1
polarization in the chain, and the strength of scattering to particular
states is determined by their overlap with this spin flip state. This means
that the multiparticle states will be sampled with particles created close
together. This is where interactions of the particle wavefunctions are most
important, making neutrons a sensitive technique for looking at overlap
effects. An interesting consequence of this is that the short range
interactions between particles can have a large influence on measured
correlation functions with little effect on the thermodynamics of the spin
chain, an effect noted for spinons in uniform chains \cite{bernevig01}.
Similarly, the thermodynamic influence of bound states at low temperatures
vanishes as $1/N$ yet they have a finite scattering cross section as
discussed below.

\subsection{Neutron scattering measurements}

We made our measurements of the inelastic scattering cross-section of CN
using the SPINS cold neutron triple-axis spectrometer at the NIST Center for
Neutron Research. The same high quality sample of copper nitrate used by Xu 
\textit{et al.}\ \cite{xu00} was utilized. This 14.1g sample consists of
four coaligned single crystals of CN, with deuterium substituting for 92\%\
of hydrogen. The substitution of D for H was made as it reduces
significantly the background from incoherent scattering of neutrons but does
not change the magnetic properties of the material. The sample was mounted
with $(h,0,l)$ as the scattering plane in a pumped $^{3}$He cryostat at a
base temperature of 300~mK. This temperature is an order of magnitude
smaller than the gap energy ($\sim 4.4$ K) and the collective quantum ground
state is almost entirely free of thermally produced magnons, with a
population of these numbering less than one per two million dimer sites.

The two-magnon scattering is expected to be weak as the neutron matrix
element to it is of $\mathcal{O}(\alpha ^{2})$\ from the ground state so the
spectrometer was set up in an open configuration to gain maximum scattered
signal, and the only collimator included in the setup was of $80^{\prime }$
between monochromator and sample. A vertically focused pyrolytic graphite
PG(002) array monochromated the incident neutrons (energy $E_{i}$,
wavevector $\mathbf{k}_{i}$) and a horizontally focused array composed of
eleven independently rotatable PG(002) blades was employed to analyze the
scattered neutrons ($E_{f}$, $\mathbf{k}_{f}$). A cooled Be filter was
placed in the incident beam before the sample to remove higher-order
contamination from the beam. The actual neutron energy transfer to the
sample being 
h{\hskip-.2em}\llap{\protect\rule[1.1ex]{.325em}{.1ex}}{\hskip.2em}%
$\omega =E_{i}-E_{f}$ and the wavevector transfer is $\mathbf{Q}=\mathbf{k}%
_{i}-\mathbf{k}_{f}$.

Measurements of scattering cross-section were made by fixing the final
energy at $E_{f}=2.5$~meV ($k_{f}=1.10$ \AA $^{-1}$)\ and scanning incident
energy $E_{i}$\ at various fixed wavenumber transfers along the chain, $k=%
\mathbf{Q}\cdot \widehat{\mathbf{d}}$ (i.e. the component of the scattered
wavevector along the important chain direction). Although there are actually
two types of chain in CN (with repeats $\mathbf{b}=[1,1,1]/2$ and $\mathbf{b}%
^{\prime }=[1,\overline{1},1]/2$) this is not important in our case - we
study the $(h,0,l)$ scattering plane where the chains give identical
contributions (see Fig. 3).

With an open scattering configuration instrumental resolution is an
important consideration. The spectrometer resolution represents the spread
in coordinate space ($\mathbf{Q},\omega $) sampled by the instrument at each
measured point. The energy resolution of the spectrometer is of Gaussian
profile with a full-width-half-maximum (FWHM) at $E_{f}=2.5$ meV and 
h{\hskip-.2em}\llap{\protect\rule[1.1ex]{.325em}{.1ex}}{\hskip.2em}%
$\omega =0.8$ meV of $\sim 0.10$ meV. The $\mathbf{Q}$\ resolution is
dominated by the wide angular acceptance ($14^{\circ }$) of the analyzer on
the scattered side - it is highly elongated along a direction within the
scattering plane that is perpendicular to the scattered wavevector - and
approximating the measured angular dependence by a Gaussian profile gives a
FWHM\ of $\sim 0.2$ \AA $^{-1}$. This resolution width is very considerable,
however by using calculated scan trajectories that maintain the final
wavevector $\mathbf{k}_{f}$ along the crystallographic $(1,0,1)$ direction,
so as to integrate over the nondispersive directions between chains, good
wavenumber resolution in $k$ along the important quantum spin chain
directions (estimated at of order $0.02$ \AA $^{-1}$) is maintained.
\begin{figure}[ht]
\includegraphics[viewport=45 150 500 550,width=3in,clip]
{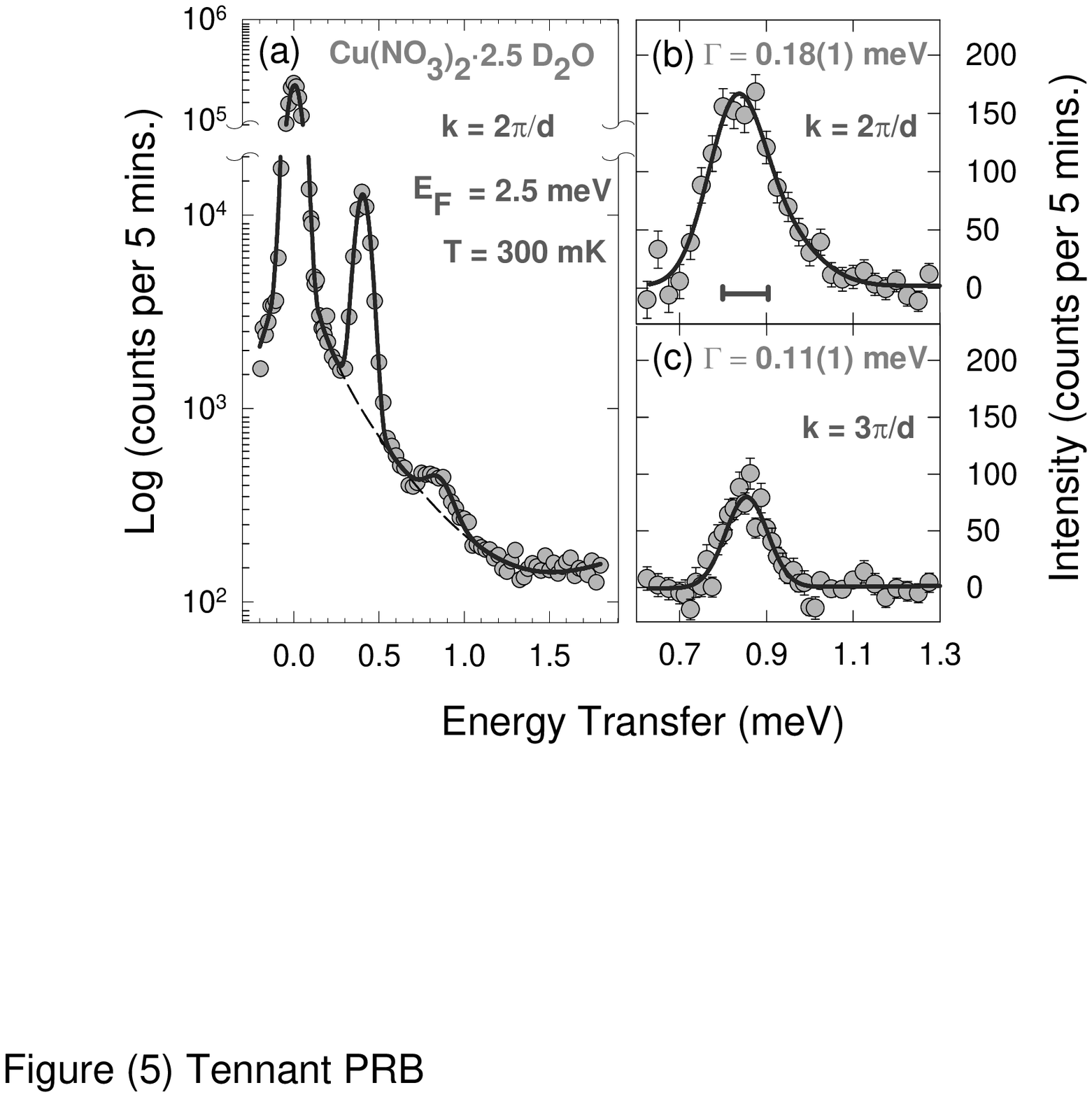}
\caption{\label{fig:num4} (a) Low temperature scattering at $k=2%
\protect\pi /d$. The dashed line is a fitted background and the solid line
is a fit to the scattering described in the text. (b) Two magnon scattering
with background subtracted off. The solid line is a fit (see text). The
solid bar indicates the instrumental resolution. (c) Two-magnon scattering
for $k=3\protect\pi /d$ with nonmagnetic background subtracted off.}
\end{figure}

\section{Results and Analysis}

Fig. 4 shows some of the results of scans in energy performed on CN: Panel
(a) shows a scan at the antiferromagnetic zone-center, $k=2\pi /d$, taken at 
$T=300$~mK. This is the wavenumber along the chain where the magnon energy
is a minimum, \textit{c.f}. Fig. 1. Strong elastic scattering from
incoherent nuclear processes is clearly seen as well as a peak at 0.4~meV as
expected for the one magnon mode \cite{xu00}, close to the dimer excitation
energy $J=0.45$~meV. A second much weaker peak appears at roughly double the
dimer energy at about 0.9~meV, which is where two-magnon scattering is
expected. On heating up the sample both the 0.4 meV and 0.9 meV peaks
disappear identifying these as being magnetic in origin.

Heating up the sample also identifies the non-magnetic background
contribution (dashed line in the figure) which consists of the incoherent
nuclear peak, modelled by a Gaussian centered at zero-energy, and a broad
contribution from thermal diffuse scattering from the analyzer which is
well-characterized by a power-times-Lorentzian (broad, quasielastic)
component decaying from zero-energy. As the background is large compared to
the two magnon signal it was studied in depth at different temperatures and
wavevectors.

Panels (b) and (c) of Fig. 4 show this peak with the modelled nonmagnetic
background subtracted at $k=2\pi /d$ and $k=3\pi /d$, respectively. This
feature is considerably weaker than the one-magnon scattering. The FWHM of
the peak narrows from 0.18(1) meV at $2\pi /d$, panel (b), to 0.11(1) meV at
the zone-boundary ($3\pi /d$), panel (c); behavior indeed consistent with
two-magnon scattering.

Fig. 1 shows the energy corresponding to a weighted average of scattering as
a thick grey line. It is notable that at $k=3\pi /d$ to a good approximation
the neutrons couple only to the bound mode, so that nearly all the
scattering weight is in it, not the continuum. The calculated neutron
scattering intensity from the bound state is $\sim 2\%$ of\ the one-magnon
intensity which agrees with the data in Fig. 4.

The one- and two-magnon scattering at 300~mK was scanned from $k=\pi /d$ to $%
5\pi /d$ in steps of $\pi /4d$. The background subtracted data are plotted
in the upper panel of Fig. 5. The lower panel shows the calculated magnetic
scattering based on the 1D perturbation theory of the Appendix with the
estimated parameters $J=0.45$ and $\alpha =0.27$. The calculation includes
the correct dimer structure factor effects and uses the true scan
trajectories in conjunction with the dimer coordinates given in Table II.
Corrections for the Cu$^{2+}$ magnetic form factor \cite{brown99} have also
been made. Instrumental line broadening has been included as well by
convolving the theoretical calculations with the estimated instrumental
resolution and sample mosaic spread. The calculation is directly comparable
with the data in the upper panel of Fig. 3, and at a qualitative level there
is good agreement with experiment.
\begin{figure}[ht]
\includegraphics[viewport=0 0 250 550,width=2.2in,clip]
{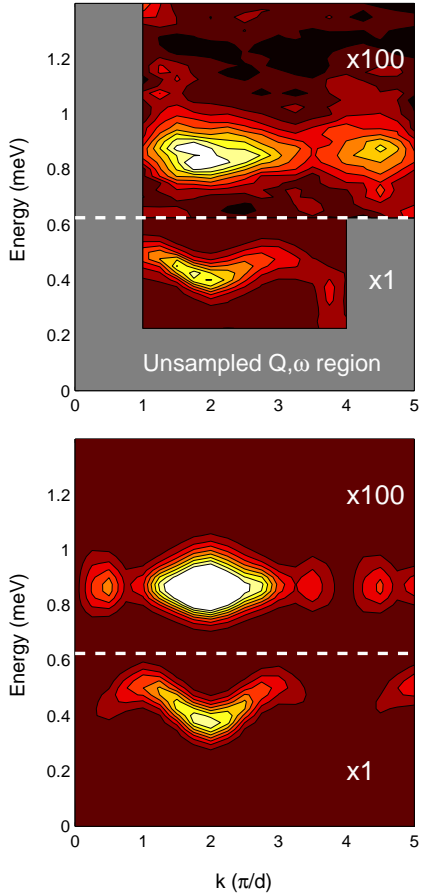}
\caption{\label{fig:num5} (Color) Upper panel shows a 
color filled contour plot of the measured
data with nonmagnetic background subtracted. Intensity is on a linear scale
indicated by color, going from dark red (minimum)\ to light yellow
(maximum). The two-magnon scattering has been enhanced by a factor of 100 to
make it visible on the same scale. Lower panel shows the calculated
scattering using perturbation theory with corrections for instrumental
resolution, multiple scattering and magnetic form factor.}
\end{figure}

Next we consider the measured one- and two-magnon scattering in more detail
and relate this to the physical picture presented by perturbation theory. A
quantitative comparison between theory and data is shown in Fig. 6. The
measured positions of one- and two-magnon peaks are plotted in the left
panel. Energies, widths and intensities for each peak were extracted by
least-squares fitting of Gaussians. In fact the wavevector and energy
resolution was not sufficiently good in this experiment to distinguish
details of line shape and the energies, widths and intensities represent the
meaningful content of the measured data. We examine the one-magnon
scattering first.

\subsubsection{\protect\bigskip One-magnon scattering}

\emph{Dispersion}: The measured dispersion of the one-magnon states is shown
in the left hand panel of Fig. 6. Considerable dispersion of the one-magnon
modes around the dimer energy (0.45 meV) is evident as expected. Although
the one magnon dispersion has been calculated to high order $\mathcal{O}%
(\alpha ^{5})$ \cite{BRT} previously, the small value of $\alpha $ in CN
means that the one-magnon dispersion should be well approximated by the $%
\mathcal{O}(\alpha )$ result, equation (2). In fact the dispersions in CN
measured by Xu \textit{et al.} \cite{xu00}\ and Stone \textit{et al.} \cite%
{stone}\ show that it is well described by%
\begin{equation}
\omega (\mathbf{Q})=J-\frac{1}{2}\sum_{\mathbf{u}}\ J_{\mathbf{u}}\cos (%
\mathbf{Q}\cdot \mathbf{u})\;
\end{equation}%
with $J=0.442(2)$ meV the dimer coupling, $J_{[111]/2}=0.106(2)$ meV the
along chain coupling, plus\ additional weak interdimer couplings $%
J_{[1/2,0,0]}^{\prime }=0.012(2)$ meV and $J_{[0,0,1/2]}^{\prime }=0.018(2)$
meV. The alternation ratio that Xu \textit{et al.}\ consider $\alpha
=J_{[111]/2}/J=0.240(5)$ is smaller than that found from the magnetization
data discussed above which is presumably due to the neglect of interchain
coupling effects in the analysis of the latter.

The solid line through the one magnon dispersion in Fig. 6 is that
calculated using the results of \cite{xu00} and it gives a reasonable
account of the data. The small discrepancies from the peak-centers measured
here are attributable due to the effects of the instrumental resolution
which averages over a large swathe of the interchain dispersion modulation.

\emph{Intensity}: The lower right-hand panel of Fig. 6 shows the one-magnon
intensities extracted from fitting. The neutron scattering matrix element $%
\mathcal{S}^{+-}(\mathbf{Q})\equiv |\langle {\Psi _{\lambda }}(k)|S_{\mathbf{%
Q}}^{+}|\Psi _{G}\rangle |^{2}$ for excitation of the one-magnon states in
the AHC to $\mathcal{O}(\alpha ^{3})$ is \cite{BRT}%
\begin{equation}
\begin{array}{c}
\mathcal{S}_{\text{1mag}}^{+-}(\mathbf{Q})=(1-\cos (\mathbf{Q}\cdot \mathbf{%
\rho }))\cdot \\ 
\left\{ 
\begin{array}{c}
\left( 1-\frac{5}{16}\alpha ^{2}-\frac{3}{32}\alpha ^{3}\right) +\left( 
\frac{1}{2}\alpha -\frac{1}{8}\alpha ^{2}-\frac{5}{192}\alpha ^{3}\right)
\cos (kd)+ \\ 
\left( \frac{3}{16}\alpha ^{2}+\frac{7}{48}\alpha ^{3}\right) \cos (2kd)+%
\frac{5}{64}\alpha ^{3}\cos (3kd)%
\end{array}%
\right\} .%
\end{array}
\label{brt_onemag}
\end{equation}%
The leading order scattering process is from the bare dimer component of the
ground state, and the $\alpha /2\cdot \cos (kd)$ component in the one magnon
structure factor arises from an $\mathcal{O}(\alpha )$ two-dimer excitation
in the ground state. The dynamical structure factor then reflects both the
composition of the ground and excited states leading to a complex wavevector
dependence.

An interesting aspect of the one-magnon intensity $\mathcal{S}_{\text{1mag}%
}^{+-}(\mathbf{Q})$ noted in \cite{BRT} is that the spin structure factor
comes in only as a $(1-\cos (\mathbf{Q\cdot \rho }))$ modulation. So where $%
\mathbf{Q\cdot \rho }=2\pi n$ ($n$ integer) the magnetic intensity should be
zero. Although this situation occurs for the measurements at the wavenumber $%
k=3.9\pi /d$, the scattering does not go to zero because of residual
intensity from secondary elastic scattering from incoherent processes. Such
residual scattering was also observed in CN by Xu \textit{et al. }\cite{xu00}%
. The solid line in the figure then is the scattering intensity predicted
using equation (4) with $J=0.45$ meV and $\alpha =0.27$ including secondary
scattering, as well as instrumental resolution and magnetic form factor. It
is seen to account very well for the observed scattering intensity. We now
consider the two-magnon scattering.

\subsubsection{Two-magnon scattering}

Integrated peak intensities, widths and positions extracted using the
fitting method described above for the two-magnon scattering are also
plotted in Fig. 6. The peak positions are plotted in the left panel of Fig.
6 (filled circles). They are nearly dispersionless at an energy of $\sim
0.86 $ meV. The peak widths are plotted in the top right-hand panel. These
are largest around $k=2\pi /d$ and $4\pi /d$ where the continuum is expected
to be widest, and are near resolution limited around $\pi /d,3\pi /d,$ and $%
5\pi /d$. Also shown is the integrated peak intensity $\int d\omega I(\omega
)$, where $I(\omega )$ is the intensity, which peaks around $k=2\pi /d$.\ 

\emph{Intensity}:\ The integrated peak intensity of the measured two magnon
scattering includes a sum over the two-particle continuum and $S=1$ bound
mode. Taking account of the density of momentum states with energy transfer,
the scattering intensities for the continuum (eqn. (17)) and bound (eqn.
(18)) states as described in the Appendix were computed. As interchain
effects are effectively integrated over in the two-magnon scattering, and
this results in line broadening rather than shifts in energy; we thus have
implicitly included interchain coupling effects in our definition of $\alpha
=0.27$ for (1). The energy integrated two-magnon intensity, $\int d\omega
I(\omega )$, (right middle panel) shows a more complicated $k$ dependence
than the one magnon scattering. The comparison with the $\mathcal{O}(\alpha
^{2})$ calculation looks qualitatively similar to the data, however it
underestimates the scattering at $k\approx 9\pi /2d$ and overestimates it at 
$2\pi /d$, which may indicate that higher order terms in the scattering
amplitude are important. It is notable that the two-magnon intensity is very
strongly dependent on the spatial arrangement of magnetic ions.

\emph{Center}: the fitted peak centers are compared with the computed
weighted center $\left\langle \omega \right\rangle =\int d\omega I(\omega
)\times \omega $ (gray band) for the $\mathcal{O}(\alpha ^{2})$\
perturbation theory in the left panel of Fig. 6. The nearly dispersionless
extracted positions (grey filled circles) are located at the calculated
weighted-average energies (grey band) replotted from Fig. 1. The fact that
the weighted center lies below the center of the continuum is a direct
result of the movement of scattering weight towards the lower boundary due
to the magnon-magnon interaction.

\emph{Width}: the peak widths obtained from the fits are shown in the top
right-hand panel. The solid line represents that calculated using the
perturbation theory. It is the sum-in-quadrature of the instrumental
resolution width in energy and the variance $\sigma $\ of the theoretical
intensity where $\sigma ^{2}$ $=\int d\omega I(\omega )\times (\omega
-\left\langle \omega \right\rangle )^{2}$. The calculation is seen to
provide a good account of the data.

\emph{Bound mode}: One of the most interesting aspects of the multiparticle
states in the AHC\ is the existence of the bound mode below the two-magnon
continuum. The predicted wavevector dependence of the intensity (see
Appendix) is 
\begin{equation*}
\begin{array}{l}
\mathcal{S}_{\text{BS}}^{+-}(\mathbf{Q})=\left( \frac{\alpha }{4}\right) ^{2}%
\left[ 1-4\cos ^{2}(kd/2)\right] \times \\ 
\text{ \ \ \ \ \ \ \ \ \ \ \ \ }\left[ \sin \left( \mathbf{Q}\cdot (\mathbf{%
\rho }+\mathbf{d})/2\right) +3\sin \left( \mathbf{Q}\cdot (\mathbf{\rho }-%
\mathbf{d})/2\right) \right] ^{2}.%
\end{array}%
\end{equation*}%
and should be visible around $k=3\pi /d$. The binding energy of the $S=1$
state, neglecting interchain coupling, is predicted to be \cite{BRT} $%
E_{B}=J\left( \frac{1}{4}\alpha -\frac{13}{32}\alpha ^{2}\right) =0.017$ meV
for CN. The scattering in CN around $k=3\pi /d$ is centered\ at $0.852\pm
.007$ meV, which gives a binding energy of $E_{B}=0.03\pm .02$ meV. In
addition the scattering is near resolution limited, as expected for a
well-defined mode. However, although the energy and intensity around $k=3\pi
/d$ lend support to binding around this bandwidth minimum, the experimental
error means this does not constitute definitive proof of the effect in CN.

\section{Discussion}

Our measurements establish the feasibility of studying weak multi-magnon
states using neutron scattering and raise a number of important issues.
Firstly the introduction of experimental data highlights the need for
practical techniques for calculating the multiparticle excitation spectra
and cross sections for realistic spin models. Our perturbation theory,
although useful for interpreting results, is of too low order to
quantitatively account for our measurements and in addition does not include
interchain coupling. Very powerful linked-cluster-expansion techniques have
recently been introduced that allow multiparticle spectra to be calculated
to high order\cite{trebst00,zheng01} for the AHC and the extension of these
to calculations of the neutron scattering cross-section would be a
significant development. Analytical approaches based on Green function
techniques may also prove fruitful. Secondly, measurement of weak
multiparticle signals preludes the access of neutron scattering to measuring
bound two-magnon states. Interactions such as next nearest neighbor coupling
should further stabilize bound modes and make measurement of these easier. A
question which requires further investigation is the stability of such bound
modes to thermal fluctuations and also interchain coupling. An alternative
route for investigating the phenomenology of particle binding is through
solitonic systems such as the 1D Ising chain with small XY-like terms \cite%
{Ishimura}. In this case the coupling of neutrons to pairs of $S=1/2$\
solitons (also called spinons) is at zeroth order and therefore strong.
Binding of solitons only occurs when extra terms are included in the
Hamiltonian, such as exchange mixing \cite{Goff}, next-nearest neighbor
coupling \cite{Matsubara}, or transverse field \cite{ghosh01}. Evidence for
this binding phenomenon in and Ising-like chain has been observed recently
in neutron scattering experiments \cite{Goff}.

Time-of-flight (TOF) neutron spectrometers give potentially much better
energy resolution than conventional triple-axis instrumentation and could
provide definitive proof of the bound state in the alternating chain by
resolving the bound mode from the continuum. Previously, TOF techniques have
proven successful in the study of similar binding effects at the bandwidth
minimum of the two-soliton continuum scattering of the $S=1/2$\ XXZ Ising
chain material CsCoCl$_{3}$ \cite{Goff}. Although limited neutron fluxes may
make such measurements difficult for CN \cite{xu00} these should be feasible
and we plan to make such measurements in the near future.

\begin{figure}[ht]
\includegraphics[viewport=75 100 520 500,width=3in,clip]
{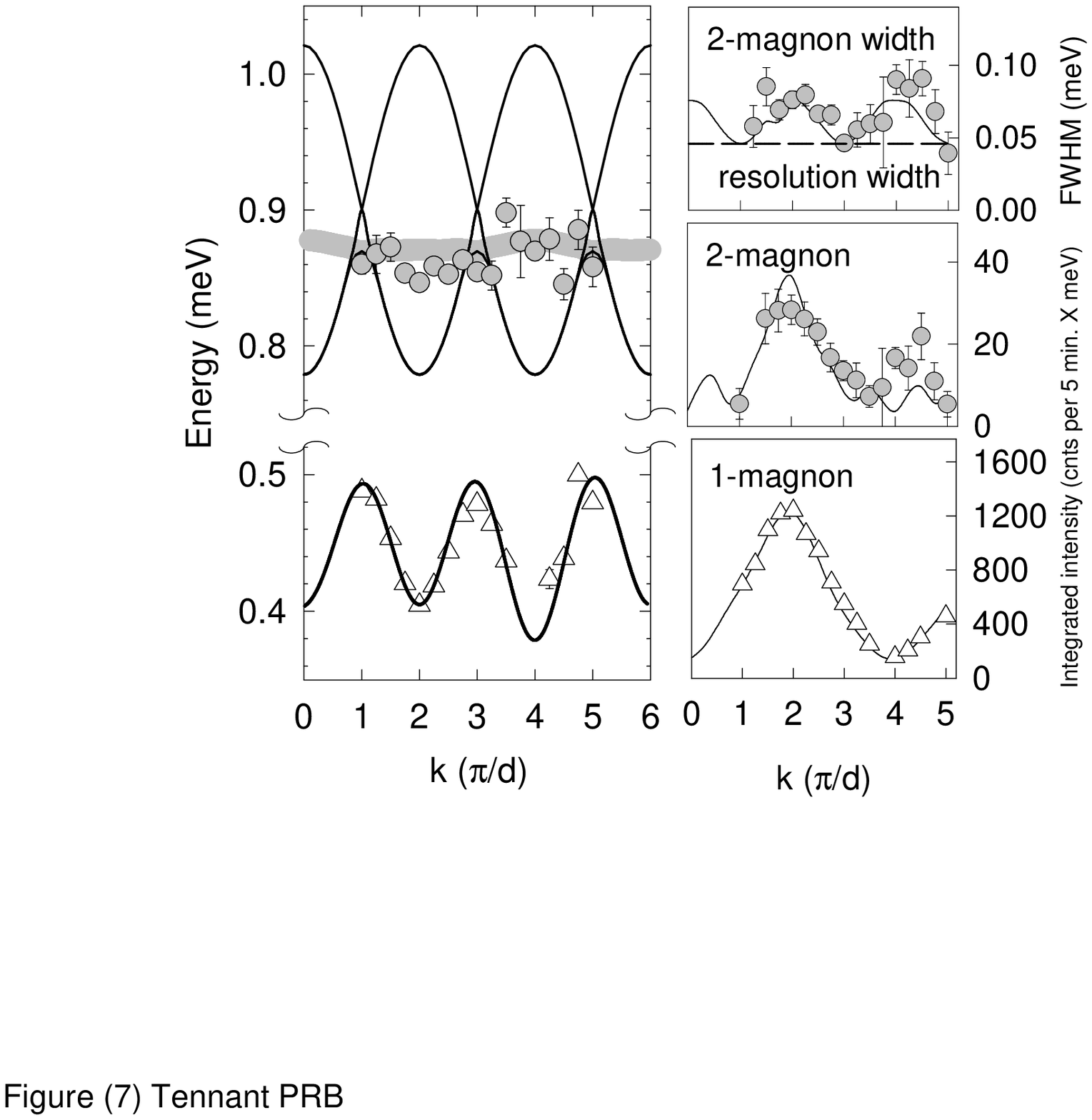}
\caption{\label{fig:num6} Comparison of theory and data.
Left panel shows fitted positions of observed sattering using perturbations
theory (see text). Right lower panel shows fitted one magnon intensity
compared with perturbation theory (see text). Right upper panel shows a
comparison of the two-magnon intensity with perturbation theory (see text).}
\end{figure}

\section{\protect\bigskip Summary}

In summary, we have used inelastic neutron scattering to investigate the
ground and excited states of the near-ideal alternating Heisenberg chain
material Cu(NO$_{3}$)$_{2}\cdot 2.5$D$_{2}$O, and also derived the
scattering analytically to lowest order in perturbation theory. Our
measurements are consistent with the predictions of this model for several
magnetic properties of this system, including the ground state energy, one-
and two-magnon excitation spectra and intensities, and possibly the
existence of a two-magnon bound state. Much experimental work remains to be
done to establish the phenomenology of binding in isotropic 1D systems.

We wish to thank Drs B. Lebech, R. Hazell, B. Lake, P-A Lindg\aa rd, and D.
McMorrow for their help and advice and also Ris\o\ National Laboratory for
generous support. This work was partly supported by Oak Ridge National
Laboratory, managed by UT-Battelle, LLC, for the US Dept. of Energy under
contract DE-AC05-00OR22725. The NSF supported work at SPINS through
DMR-9423101 and work at JHU through DMR-9453362 and DMR-9801742. DHR
acknowledges the generous support of the David and Lucile Packard
Foundation.Work at Harvard (KD) was supported by NSF-DMR grants 9981283,
9714725, and 9976621

\section{\protect\appendix Appendix}

\subsection{Ground state}

Here we apply the perturbation theory of \cite{BRT} to two-magnon
scattering. The single dimer eigenstates (labelled $\text{G},1,0,\overline{1}
$) are listed in Table I and the ground state of the uncoupled dimers ($%
\alpha =0$)\ is a direct product of dimer ground states 
\begin{equation}
\Psi _{0}=\prod_{m=1}^{N_{d}}|G_{m}\rangle ,\ E_{0}=-\frac{3JN_{d}}{4}.
\end{equation}%
As total spin $S_{T}=\sum_{m=1}^{N_{d}}S_{m}$, and $S_{T}^{z}=%
\sum_{m=1}^{N_{d}}S_{m}^{z}$ are constants-of-the-motion for the Hamiltonian 
$H$ they organize the Hilbert space. For notation we introduce dimer
creation operators $a_{m}^{+}|G_{m}\rangle =|a_{m}\rangle $ where $a=1,0,%
\overline{1}$ label excited dimer states. In real space we denote the singly
excited states with quantum numbers $(S_{T},S_{T}^{z})=(1,a)$ as $|a\rangle
_{m}=a_{m}^{+}\Psi _{0}$ and the doubly excited with $|(S_{T},S_{T}^{z})%
\rangle _{m,\nu }$ as the Clebsh-Gordan combinations of excitations at dimer
sites $m$ and $m+\nu $ \textit{i.e.} $|(0,0)\rangle _{m,\nu }=1/\sqrt{3}%
\{1_{m}^{+}\overline{1}_{m+\nu }^{+}-0_{m}^{+}0_{m+\nu }^{+}+\overline{1}%
_{m}^{+}1_{m+\nu }^{+}\}\Psi _{0}$, $|(1,1)\rangle _{m,\nu }=1/\sqrt{2}%
\{0_{m}^{+}1_{m+\nu }^{+}-1_{m}^{+}0_{m+\nu }^{+}\}\Psi _{0},$ \textit{etc}.

As the alternating chain has translational symmetry plane-wave states prove
convenient 
\begin{equation}
|a\rangle _{k}\equiv \frac{1}{\sqrt{N_{d}}}\sum_{m=1}^{N_{d}}e^{imkd}|a%
\rangle _{m},
\end{equation}%
and 
\begin{equation}
|(S,S^{z})\rangle _{k,\nu }\equiv \frac{1}{\sqrt{N_{d}}}%
\sum_{m=1}^{N_{d}}e^{imkd}|(S,S^{z})\rangle _{m,\nu }.
\end{equation}%
where the allowed momenta are $k_{n}=2n\pi /N_{d}d$, where $n$ are integers
from $-N_{d}/2$ to $N_{d}/2$. The action of the Hamiltonian on the basis
states has been considered in \cite{BRT} and the ground state to $\mathcal{O}%
(\alpha )$ is 
\begin{equation}
\Psi _{G}=\eta _{0}[\Psi _{0}-\alpha \frac{\sqrt{3}}{8}%
\sum_{m=1}^{N_{d}}|(0,0)\rangle _{m,1}]
\end{equation}%
where $\eta _{0}=1-(3/128)\alpha ^{2}N_{d}$.

Neutron scattering measures the square of the expectation value of the spin
operator $S^{+}(\mathbf{Q})=\left( 2N_{d}\right)
^{-1/2}\sum_{m=1}^{N_{d}}\sum_{p=\pm }\exp (i\mathbf{Q}\cdot \mathbf{r}%
_{m,p})S_{m,p}^{+}$ between eigenstates, where $\mathbf{Q}$ is the
wavevector transfer of the neutron. The action of this spin operator
applied\ to the ground state is%
\begin{equation}
\begin{array}{l}
S^{+}(\mathbf{Q})\left| \Psi _{G}\right\rangle =A_{\mathbf{Q}}|1\rangle
_{k}+B_{\mathbf{Q}}|(1,1)\rangle _{k,1}, \\ 
A_{\mathbf{Q}}=\frac{1}{\sqrt{2}}\left( e^{i\mathbf{Q}\cdot \mathbf{\rho }%
/2}-e^{-i\mathbf{Q}\cdot \mathbf{\rho }/2}\right) \left( 1+\frac{\alpha }{4}%
\cos (kd)\right) , \\ 
B_{\mathbf{Q}}=\frac{\alpha }{8}\left( e^{i\mathbf{Q}\cdot \mathbf{\rho }%
/2}+e^{-i\mathbf{Q}\cdot \mathbf{\rho }/2}\right) \left( e^{ikd}-1\right) .%
\end{array}%
\end{equation}%
where $k=\mathbf{Q}\cdot \widehat{\mathbf{d}}$\ is the wavenumber of the
states excited by this operator. The neutron scattering matrix element at $%
T=0$ to state $\lambda $, is given by $\mathcal{S}_{\lambda }^{+-}(\mathbf{Q}%
)=\left| \left\langle \Psi _{\lambda }\left| S_{\mathbf{Q}}^{+}\right| \Psi
_{G}\right\rangle \right| ^{2}$.

\subsection{One-magnon states}

The one-magnon wavefunctions to order $\mathcal{O}(\alpha )$ are \cite{BRT}%
\begin{eqnarray}
\Psi _{\text{1mag}} &=&|a\rangle _{k}+\frac{\alpha }{2\sqrt{2}}%
(e^{ikd}+1)|(1,a)\rangle _{k,1}- \\
&&\frac{\alpha \sqrt{3}}{8\sqrt{N_{d}}}\sum_{m,m^{\prime }=1}^{N_{d}}\delta
_{m\neq m^{\prime }}e^{imkd}a_{m}^{+}|(0,0)\rangle _{m^{\prime },1}  \notag
\end{eqnarray}%
and form an $S=1$ triplet with an energy gap $\Delta =J-\alpha J/2$ above
the ground state and dispersion 
\begin{equation}
\omega _{1\text{mag}}(k)=J-\alpha J/2\cos (kd).
\end{equation}%
Application of the $S^{+}(\mathbf{Q})$ operator gives the intensity to $%
\mathcal{O}(\alpha )$%
\begin{equation*}
\begin{array}{c}
\mathcal{S}_{\text{1mag}}^{+-}(\mathbf{Q})=(1-\cos (\mathbf{Q}\cdot \mathbf{%
\rho }))\cdot \\ 
\left\{ 1+\frac{1}{2}\alpha \cos (kd)\right\} .%
\end{array}%
\end{equation*}

\subsection{Two-magnon states}

Ignoring states higher than two-excited-dimer (as they do not contribute to
the neutron-scattering matrix element to lowest order in $\alpha $), the
action of the Hamiltonian is \cite{BRT} 
\begin{equation*}
H|a\rangle _{k}=\alpha _{k}|a\rangle _{k}+\sqrt{2}\gamma _{k}^{\dagger
}|(1,a)\rangle _{k,1}
\end{equation*}%
and 
\begin{eqnarray}
&&H|(1,a)\rangle _{k,\nu }=  \notag \\
&&\left\{ 
\begin{array}{l}
\sqrt{2}\gamma _{k}|a\rangle _{k}+(\beta -\epsilon )|(1,a)\rangle
_{k,1}+\gamma _{k}|(1,a)\rangle _{k,2},\;\nu =1 \\ 
\beta |(1,a)\rangle _{k,\nu }+\gamma _{k}|(1,a)\rangle _{k,\nu +1}+\gamma
_{k}^{\dagger }|(1,a)\rangle _{k,\nu -1},\text{ }\nu >1%
\end{array}%
\right.
\end{eqnarray}%
where $\alpha _{k}=J-(\alpha J/2)\cdot \cos (kd)$, $\beta =2J$, $\epsilon
=\alpha J/4$, $\gamma _{k}=-\frac{\alpha J}{4}\left( 1+e^{-ikd}\right) $,
and $\dagger $ denotes complex conjugation. The excitation spectrum can be
calculated by direct diagonalization of a large number of dimers and
application of the matrix element above or by analytical solution.

Approximate analytical wavefunctions for the $S=1$ states can be calculated
using elementary scattering theory, see \cite{Matsubara}. Ignoring the
coupling to the one-excited-dimer states for the time being, the two-magnon
wavefunctions are%
\begin{equation}
\Psi _{\text{2mag}}(k)=\sum_{\nu =1}^{N_{d}-1}b_{\nu }\exp (i\theta \nu
)|(1,1)\rangle _{k,\nu }.
\end{equation}%
where for $\theta =-i\log (\sqrt{\gamma _{k}^{\dagger }/\gamma _{k}})=kd/2$
the time-independent Schr{\"{o}}dinger equation reduces to solving the real
and symmetric system of equations 
\begin{equation}
\begin{array}{l}
\lambda b_{1}=(\beta -\epsilon )b_{1}+\widetilde{\gamma }_{k}b_{2} \\ 
\vdots \\ 
\lambda b_{\nu }=\beta b_{\nu }+\widetilde{\gamma }_{k}(b_{\nu +1}+b_{\nu
-1}) \\ 
\vdots \\ 
\lambda b_{N_{d}-1}=(\beta -\epsilon )b_{N_{d}-1}+\widetilde{\gamma }%
_{k}b_{N_{d}-2}%
\end{array}%
\end{equation}%
where $\widetilde{\gamma }_{k}=\sqrt{\gamma _{k}^{\dagger }\gamma _{k}}%
=\alpha J/2\cdot |\cos (kd/2)|$ and the term $\exp (i\theta \nu )$ serves to
transform to the center-of-momemtum (center $R$) frame where magnons are at $%
r_{i}=R-\nu /2$ and $r_{j}=R+\nu /2$ with total momentum $K=k_{1}+k_{2}=kd$.

\subsubsection{Magnon-pair-state solutions}

Magnon-pair-states comprise particles that are free at large distances and
for particle conservation in one-dimension the interactions introduce a
phase factor $\phi $ on scattering. This state corresponds to 
\begin{equation}
b_{\nu }^{\mu }=X_{0}\left( \exp \left( ip_{\mu }\nu \right) -\exp \left(
-i\left( p_{\mu }\nu -\phi _{\mu }\right) \right) \right)
\end{equation}%
where the normalization constant $X_{0}\simeq 1/\sqrt{N_{d}}$, $p_{\mu }$ is
the relative momentum, and $\mu =1,2,...,N_{d}-1$ index the eigenstates. The
phases and momenta are determined by the boundary conditions of the
particles and their interaction energy. The standard method of solving for
these boundary conditions \cite{Matsubara,Mattis} is to introduce the single
site coefficients $b_{0}$ and $b_{N}$ and set $\epsilon b_{1}=\widetilde{%
\gamma }_{k}b_{0}$ and $\epsilon b_{N_{d}-1}=\widetilde{\gamma }%
_{k}b_{N_{d}} $ and substitute into equation (14). The momentum $p_{\mu }$
and phase $\phi _{\mu }$ solve for the constraints on $b_{0}$ and $b_{N_{d}}$
when%
\begin{equation}
e^{i\phi _{\mu }}=\frac{\widetilde{\gamma }_{k}+\epsilon \exp \left( -i\frac{%
\pi \mu +\phi _{\mu }}{N_{d}}\right) }{\widetilde{\gamma }_{k}+\epsilon \exp
\left( i\frac{\pi \mu +\phi _{\mu }}{N_{d}}\right) },p_{\mu }=\frac{\pi \mu
+\phi _{\mu }}{N_{d}}
\end{equation}%
and the eigenvalues are $\lambda _{\mu }=2J-\alpha J\cos (kd)\cos (p_{\mu })+%
\mathcal{O}(1/N_{d})$. In the time-dependent Schr\"{o}dinger picture this
state corresponds to two particles with wavepackets at positions $%
r_{i}=R+\nu /2$ and $r_{j}=R-\nu /2$ and with momenta $k_{1}=(K+p)/2$ and $%
k_{2}=(K-p)/2$ that scatter via the $\mathbf{S}$-matrix $\mathbf{S}%
_{k_{1},k_{2}}=-\exp (i\phi _{k_{1}-k_{2}})$, and the eigenspectrum is 
\begin{equation}
\omega _{k_{1},k_{2}}=\omega _{\text{1mag}}(k_{1})+\omega _{\text{1mag}%
}(k_{2})
\end{equation}%
with $\omega _{\text{1mag}}(k)=J-\alpha J/2\cdot \cos (kd)$ as above.

Including the coupling to the $|1\rangle _{k}$ states as a perturbation
gives the approximate wavefunction%
\begin{equation}
\Psi _{2\text{mag}}(k,\mu )=-\frac{\alpha c_{1}^{\mu }}{2\sqrt{2}}%
(1+e^{-ikd})|1\rangle _{k}+\sum_{\nu =1}^{N_{d}-1}c_{\nu }^{\mu
}|(1,1)\rangle _{k,\nu }
\end{equation}%
where $c_{\nu }^{\mu }=b_{\nu }^{\mu }\exp \left( ikd\nu /2\right) $. The
neutron scattering matrix element can then be computed straightforwardly for
a large $N_{d}$ system by evaluating the closure with equation (8).

\subsubsection{Bound state solutions}

When $\widetilde{\gamma }_{k}<\epsilon $, $\phi _{\mu }$ cannot be solved
with $\mu =1$ and $N_{d}-1$, and the $c_{\nu }^{1}$ and $c_{\nu }^{N_{d}-1}$
solutions comprise exponentially decaying bound states solutions below the
two magnon continuum. The $S=1$ bound state wavefunctions and energy have
previously been given by \"{U}hrig and Schulz \cite{uhrig}, and Damle and
Nagler \cite{Damle}. It is specified by equation (18) with%
\begin{equation}
b_{\nu }^{1}=\sqrt{\frac{\epsilon ^{2}-\widetilde{\gamma }_{k}^{2}}{%
\widetilde{\gamma }_{k}^{2}}}\exp (-\kappa \nu ),
\end{equation}%
where $\exp (-\kappa )=-\widetilde{\gamma }_{k}/\epsilon $ which has
dispersion energy 
\begin{equation*}
\omega _{\text{BS}}=2J-\frac{\alpha J}{4}\left( 4\cos ^{2}(kd/2)+1\right) .
\end{equation*}%
An interesting feature of the bound state solution is that it only exists
over the range $\left| n\pi -kd\right| \leqslant \pi /3$, where $n$ is an
odd integer, and so there is an $S=1$ bound mode only for small wavevectors
around the narrowest part of the continuum. Using our wavefunction we
calculate the neutron scattering strength from the bound state to be%
\begin{equation*}
\begin{array}{l}
\mathcal{S}_{\text{BS}}^{+-}(\mathbf{Q})=\left( \frac{\alpha }{4}\right) ^{2}%
\left[ 1-4\cos ^{2}(kd/2)\right] \times \\ 
\text{ \ \ \ \ \ \ \ \ \ \ \ \ }\left[ \sin \left( \mathbf{Q}\cdot (\mathbf{%
\rho }+\mathbf{d})/2\right) +3\sin \left( \mathbf{Q}\cdot (\mathbf{\rho }-%
\mathbf{d})/2\right) \right] ^{2}.%
\end{array}%
\end{equation*}


\begin{thebibliography}{99}
\bibitem{space_group} Note that the space group for Cu(NO$_{3}$)$_{2}\cdot
2.5$H$_{2}$O has been incorrectly given in previous work as I 1 2/a 1.
\end{thebibliography}
\end{document}